# VLA Observations of the Gravitational Lens System Q2237+0305

E.E. Falco, J. Lehár
Center for Astrophysics, 60 Garden St, Cambridge, MA 02138
efalco@cfa.harvard.edu, jlehar@cfa.harvard.edu

R.A. Perley
National Radio Astronomy Observatory, P.O. Box 0, Socorro, NM 87801
rperley@nrao.edu

J. Wambsganss
Astrophysikalisches Institut Potsdam, Germany
jwambsganss@aip.de

and

M.V. Gorenstein
Waters Corporation, 34 Maple St, Milford, MA 01757
gorenstein_marc@waters.com

## ABSTRACT

We report observations of the four-image gravitational lens system Q2237+0305 with the VLA at 20 cm and 3.6 cm. The quasar was detected at both frequencies ($\approx 0.7$ mJy) with a flat spectrum. All four lensed images are clearly resolved at 3.6 cm, and the agreement of the radio and optical image positions is excellent. No radio emission is detected from the lensing galaxy, and any fifth lensed quasar image must be fainter than $\sim 20\%$ of the A image flux density. Since the optical quasar images are variable and susceptible to extinction, radio flux ratios provide the best measurement of the macrolensing magnification ratios. The radio B/A and C/A image flux ratios are consistent with the observed range of optical variations, but the D/A ratio is consistently higher in the radio than in the optical. The radio ratios are consistent with magnification ratios predicted by lens models, and weaken alternative interpretations for Q2237+0305. More accurate radio ratios can distinguish between the models, as well as improve our understanding of both microlensing and extinction in this system.

*Subject headings:* Gravitational Lenses — Dark Matter — Quasars: individual (Q2237+0305)

# 1. Introduction

Among the surprises from the CfA redshift survey was the discovery of the gravitational lens Q2237+0305 (Huchra *et al.* 1985). In this system, the light from a distant quasar, at $z \sim 1.7$, is split into four images by the gravitational field of a nearby spiral galaxy, at $z \sim 0.04$. The images of the quasar are symmetrically arranged around the nucleus of the lensing galaxy, with a maximum separation of $\sim 1\rlap{.}''8$ (Yee 1988). Gravitational lensing in such systems can be an important astrophysical tool. For example, observations of the four images can be used to constrain the distribution of dark matter in the foreground galaxy. Also, if time delays can be measured between the quasar images, they can be used with the lens model to estimate the lens distance, independent of standard distance measures, providing a new estimate of $H_0$. Since the lensing galaxy in Q2237+0305 is very nearby, its properties can be studied in much greater detail than for any other lensed system.

A very satisfying lensing interpretation for the Q2237+0305 system soon developed on the basis of many optical observations. Although early studies of the quasar images were complicated by the presence of the bright foreground galaxy (e.g., Yee 1988; Racine 1991), contemporary lens models easily reproduced the observed quasar image positions (e.g., Schneider *et al.* 1988; Kent & Falco 1988; Kochanek 1991). The impressive Hubble Space Telescope results (Crane *et al.* 1991; Rix *et al.* 1992) led to improved lens models which reproduced the image positions to $<0\rlap{.}''02$ (e.g., Rix *et al.* 1992; Wambsganss & Paczyński 1994, hereafter WP94).

The optical quasar image brightnesses, however, proved difficult to interpret. The images were found to be variable (Irwin *et al.* 1989). The expected image delays in Q2237+0305 are only a few hours, so if the variations are intrinsic, the image light curves should be identical on observed time scales. But the observed fluctuations are uncorrelated (Irwin *et al.* 1989; Corrigan *et al.* 1991; Racine 1992; Houde & Racine 1994; Lewis *et al.* 1996). This is probably due to "microlensing", where individual stars in the lensing galaxy affect the image magnifications (Chang & Refsdal 1979; Wambsganss *et al.* 1990). Microlensing depends strongly on the source size, with smaller sources giving rise to stronger and more rapid fluctuations. Furthermore, extinction by dust in the lens is observed in Q2237+0305 (Nadeau *et al.* 1991), and could also affect the image brightnesses.

Radio observations are more likely to determine the macrolensing image magnifications. The emission region in quasars is believed to be much larger in the radio than in the optical (e.g., Antonucci 1993), so microlensing is probably not an issue (e.g., Young 1981). Extinction from the galaxy is also not likely to be important at centimeter wavelengths. Unfortunately, the large size of the radio source also suggests that unless it contains a highly-beamed jet, the radio source is not likely to vary on timescales comparable to the expected image delays. However, a determination of the "macromagnification" ratios for the lensed images is very useful to refine lens models, and for further studies of microlensing effects in Q2237+0305.

## 2. Observations

The radio counterpart to Q2237+0305 was first detected on 1985.04.18, with a 3 hr observation at L band (20 cm, see Figure 1), using the NRAO Very Large Array in the B configuration. We observed the target in two 50 MHz bandpasses centered on 1.465 GHz and 1.515 GHz. The data were calibrated and mapped in the AIPS software package following standard procedures (Fomalont & Perley 1989), with the flux densities calibrated to 3C 286 on the Baars *et al.* (1977) scale. The FWHM angular resolution was $\Theta \approx 4''$, and the off-source map rms was $56\,\mu$Jy/beam, consistent with the expected thermal noise. A faint radio source was detected at right ascension $\alpha$=22:37:57.72 and declination $\delta$=+03:05:50.3 (B1950), shown in Figure 1. We estimate a positional uncertainty of $\sim 0''\!.23$ ($1\sigma$), by combining a $\sim 0''\!.2$ VLA astrometric error (e.g., Lawrence *et al.* 1986) with signal-to-noise considerations (Fomalont 1989). The integrated flux density was $0.832 \pm 0.087$ mJy, and the source was unresolved. VLA observations at higher frequencies could resolve the lensed images of Q2237+0305, but only with much longer integrations.

We observed Q2237+0305 on 1995.06.25 for eleven hours with the VLA at X band (3.6 cm), in the A configuration. The center frequencies of the two 50 MHz bandpasses were 8.415 GHz and 8.465 GHz. Twenty-six antennas were active for most of the run, although two of these failed towards the end. We interleaved observations of the target and a nearby VLA phase calibrator (B 2210+016) at 15 min intervals, and observed two flux calibrators (3C 286 and 3C 48) at the ends of the run.

The data were edited and calibrated using standard VLA procedures (Fomalont & Perley 1989) in the AIPS software package. The flux densities were calibrated to 3C 286 and 3C 48, again using Baars *et al.* (1977). Since Q2237+0305 is too faint for self-calibration at X band, the antenna gains were interpolated from the phase calibrator solutions. The antenna phases were unstable for observations at elevations < 30 deg. There were also large phase fluctuations affecting the far antennas during the first half of the run, which were probably due to weather ($\sim$50% cloud cover, which cleared in several hours). So we analyzed both the full 11 hr set of data, and a restricted set with stable calibrator phases. The restricted set uses all of the available antennas for $\sim$4.9 hours, and amounts to about half of the data.

We mapped the source using CLEAN deconvolution in the AIPS task MX. Initial wide-field maps showed no other radio sources in the VLA primary beam, so we limited subsequent mapping and CLEANing to a $6''.4$ square region centered on Q2237+0305. The final maps for the full and restricted data are shown in Figure 2. For both maps, the FWHM angular resolution is $\Theta \sim 0''.3$ (about 6 pixels), and the off-source rms is consistent with the expected thermal noise. In the restricted map, the rms noise is $9.8\,\mu$Jy/beam, and the total flux density for Q2237+0305, within a $2''.8$ square aperture, is $593 \pm 88\,\mu$Jy. This leads to a flat radio spectral index of $\alpha_{XL} = -0.18 \pm 0.10$, assuming no substantial variations since 1985. The source appears to be $\sim 25\%$ fainter in the map from the full data set, probably due to atmospheric phase decorrelation.

The maps show four components, named following Yee (1988), but the C component identification is ambiguous in the map from the full data. When we mapped subsections of the restricted data, the component shapes changed substantially, and there were no significant flux density variations. So the apparent component shapes are probably due to antenna phase errors. The A component is at $\alpha$=22:37:57.754, $\delta$=+03:05:49.45 (B1950), with an absolute position error of $<0''.2$ (e.g., Lawrence et al. 1986), and the peak of the L band source falls between the three brightest X band components, as expected. Table 1 presents the component positions and flux densities. The flux densities were integrated using a square aperture whose size was chosen to maximize the component signal/noise ratios, and to enclose all the scattered signal. An aperture was also positioned at the center of the lensing galaxy G, using the optical offset (Crane et al. 1991) from the B component.

## 3. Discussion

All four of the lensed quasar images are detected at X band. The published absolute coordinates of the optical source (Huchra et al. 1985) are only accurate to $\sim 10''$, so we confirmed the identification using the HST Digitized Sky Survey. The center of the lensing galaxy G is at $\alpha$=22:37:57.76, $\delta$=+03:05:50.6 (B1950), with a $\sim 1''$ uncertainty, consistent with the radio position. The agreement of the radio and optical relative positions for the four lensed images is also excellent. Allowing for a constant shift, the $\Delta\alpha$ and $\Delta\delta$ residuals between the radio (Table 1) and the optical components (Crane et al. 1991, Table 2) are within the measurement uncertainties (rms$\sim 0''.02$). Note that the D image is much more prominent in the radio, and that no radio counterpart is detected for the lensing galaxy. If there is a fifth image, it should be close to the core of the lensing galaxy. So the absence of emission near G also sets a $1\sigma$ upper limit of $\sim 10\,\mu$Jy/beam, for a fifth quasar image, or about one fifth of the A image flux density.

Unlike the optical image brightnesses, the radio component flux densities agree with lens model predictions (see Figure 3). For example, all of the existing models which had been constrained only on positions predict a D/A image brightness ratio $R_{\rm DA} \approx 1$, but all of the optical observations to date have found $R_{\rm DA} < 0.6$. The radio ratios are much more consistent with the predictions, and offer the possibility of distinguishing between them. The multiparametric models of WP94 do best at reproducing the image positions, and range from roughly flat ($\beta \approx 2$) through isothermal ($\beta = 1$) to steep (point-like, $\beta = 0$) surface density profiles. The WP94 models with steep profiles are slightly more consistent with the radio $R_{\rm DA}$.

Using the "lensmod" software (see Lehár et al. 1993), we fitted lens models similar to those in WP94, and used both the optical positions (Crane et al. 1991) and our radio flux ratios as constraints. As in WP94, the non-spherical component of the mass distribution is accounted for using an external shear. The resulting reduced $\chi^2$ rose from 2.4 to 2.6 for $\beta$ increasing from 0 to 2, with steeper profiles being slightly preferred. We also considered models where the external shear was replaced with an elliptical lensing potential (Blandford & Kochanek 1987). This produced a better overall fit ($\chi^2 = 2.0$), but strongly excluded profiles with $\beta < 0.7$. This effect was even stronger when the optical positions were replaced by those of Rix et al. (1992), and it appears that an "internal" ellipticity may be superior to an external shear for modelling the lensing mass distribution (see Kochanek 1991). Finally, if the surface density profile is non-singular at the core, our upper limit on the fifth image flux density can be used to constrain a core radius parameter $\theta_c$ (Blandford & Kochanek 1987). We found that flat profiles with large core radii require a detectable fifth image. For example, the "best fit" model of WP94 ($\beta = 1.71$) with $\theta_c > 0\rlap.{''}5$ predicts a fifth image flux density exceeding $15\,\mu{\rm Jy}$. So although we cannot definitively rule out any lens models under serious consideration, the flux ratios do yield suggestive preferences.

Whether microlensing is important in the radio is determined by the size of the radio emitting region. With a flux density of $\sim 1\,{\rm mJy}$, the background source is a "radio-quiet" quasar ("RQQ", e.g., Kellermann et al. 1989; Miller et al. 1993). Size limits for the radio source can be obtained from the time scale of radio variations and from the brightness temperature limit due to Compton self-absorption (Kellerman & Pauliny-Toth 1969). A recent study of RQQ radio emission (Barvainis et al. 1996) suggests that most RQQs with flat or inverted spectra vary by a factor of two, on time scales of 3-10 years. With a $(1+z)$ time dilation factor, this implies that the source may be smaller than 0.3 light-years ($< 15\,\mu$as at $z = 1.7$, assuming $H_0 = 75$ and $q_0 = 0.5$). We know of no studies which address RQQ variability on shorter time scales. A radio source whose flux density is $1\,{\rm mJy}$ at $6\,{\rm cm}$, with a lensing magnification of $\sim 10$, and with $T_b < 10^{12}\,{\rm K}$, must exceed $\sim 1\,\mu$as in angular extent ($\sim 1000\,{\rm AU}$ at $z = 1.7$). At $z = 1.7$, the projected critical radius of solar-mass microlensing stars in the foreground galaxy is $\sim 5\,\mu$as (e.g., Wambsganss *et al.* 1990). So we cannot rule out microlensing as an important effect in the radio. Although we think radio microlensing is unlikely, this issue can only be settled by monitoring Q2237+0305 in the radio.

If the radio flux densities reflect the macrolensing magnifications, they can be used to better understand optical microlensing in Q2237+0305. A characteristic of microlensing is that high-magnification events are not randomly distributed in time, but are highly clustered (e.g., Wambsganss *et al.* 1992). Periods of rapid variability, as the source passes through bunches of caustics, are separated by quiescent intervals where the image is dimmed relative to the macromagnification. While the time scale for a single event ($\sim$months) is related to the size of the background source, the length of the quiescent periods (up to tens of years) depends on how the intervening stars are distributed, and on the macrolensing shear. Gravitational lensing conserves surface brightness, so the time-averaged optical magnifications should eventually converge to the radio values. The optical variations are very poorly sampled in time, so we roughly estimate the average ratios for each image by taking the full range of observed $R$ filter ratios (Houde & Racine 1994) and averaging the extremes. These range centers for $R_{BA}$ and $R_{CA}$ are not far from the radio brightnesses (see Figure 3). $R_{DA}$ is consistently lower in the optical, and the optical D image could well be in a quiescent interval (Lewis *et al.* 1996). In principle, the duration of the D image depression could be used to constrain the stellar clumping in the galaxy, but only with a much better understanding of both macro- and microlensing in Q2237+0305 than is presently available.

Extinction by dust in the lensing galaxy can also affect the optical ratios, and is another possible explanation for the persistently low $R_{DA}$. Nadeau *et al.* (1991) measured extinction in Q2237+0305, and could plausibly account for the optical ratios. Without reliable estimates of the macroimage magnifications, however, they had to assume that Q2237+0305 and our Galaxy have the same reddening law. This assumption led to predicted magnification ratios of $R_{AC}R_{BC}=1.2$ and $R_{AD}R_{BD}=2.4$. Our radio observations yield quite different ratios $R_{AC}R_{BC}=3.6 \pm 2.6$ and $R_{AD}R_{BD}=1.6 \pm 0.9$, but the errors are too large to exclude the Nadeau *et al.* (1991) assumption. For completeness, we repeated the extinction law derivation, using our radio ratios for the magnifications. Using the radio $R_{AD}R_{BD}$, extinction in the galaxy decreases with increasing wavelength, and is consistent with that in our Galaxy (see Nadeau *et al.* 1991, Figure 7). The radio $R_{AC}R_{BC}$, on the other hand, yields an extinction law which rises with increasing wavelength, and the C image magnification must be increased relative to A and B (by about $1\sigma$ in $R_{CA}$) to restore a reddening trend. Note, however, that these results are of very low significance. Also, the Nadeau *et al.* (1991) analysis relies on optical data taken at different times, and optical microlensing (which could be wavelength-dependent) needs to be accounted for. Moreover, extinction alone cannot explain why the amplitude of the variations is smaller in the D image than in the others (Lewis *et al.* 1996). Nevertheless, detailed studies on Q2237+0305 may yet make Q2237+0305 the most distant galaxy with a measured reddening law.

Radio observations are a promising new tool in the study of the Q2237+0305 system. Using only optical telescopes, several decades of observations would be required to establish the macromagnification ratios. Radio observations, on the other hand, should be able to break the macrolensing model degeneracy (e.g., WP94) with only a moderate increase in sensitivity. Radio maps also offer the best hope of detecting any fifth image, since there is no significant contamination from the foreground galaxy. Longer radio integrations will allow us to better understand microlensing and extinction in the system, and to test for radio variability. Since the microlensing light curves are quite different for various macrolensing models (see WP94), comparing optical microlensing to more sensitive radio ratios may also provide more constraints to the macrolensing model. Finally, seeing that the four radio components coincide exactly with the optical quasar images, the agreement of the radio ratios with lensing predictions weakens alternative interpretations for the Q2237+0305 system (e.g., Arp & Crane 1992).

It is a pleasure to thank Karl Menten and Mark Reid for useful discussions. We also thank the anonymous reviewer for helpful suggestions. The NRAO is operated by Associated Universities, Inc., under cooperative agreement with the National Science Foundation. The optical astrometry presented here was based on photographic data obtained using The UK Schmidt Telescope. The UK Schmidt Telescope was operated by the Royal Observatory Edinburgh, with funding from the UK Science and Engineering Research Council, until 1988 June, and thereafter by the Anglo-Australian Observatory. Original plate material is copyright (c) the Royal Observatory Edinburgh and the Anglo-Australian Observatory. The plates were processed into the present compressed digital form with their permission. The Digitized Sky Survey was produced at the Space Telescope Science Institute under US Government grant NAG W-2166. JL gratefully acknowleges the support of NSF grant AST93-03527.

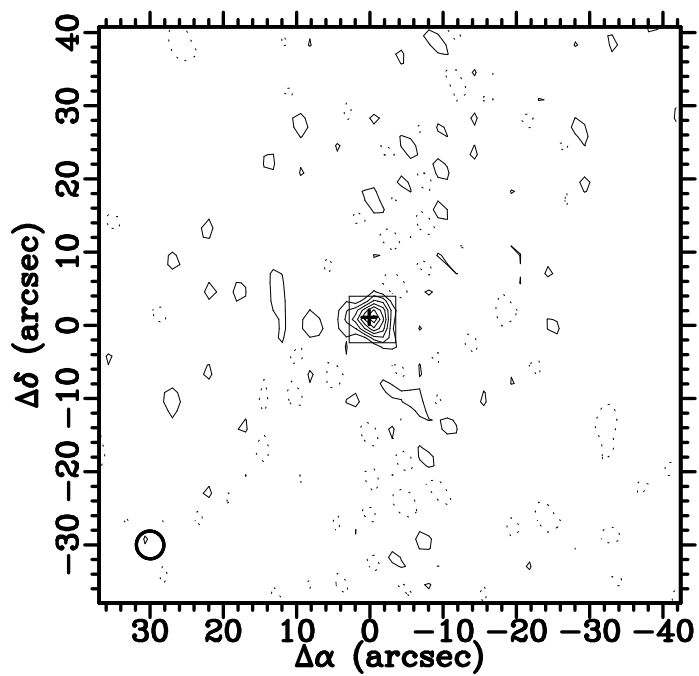

Fig. 1.— VLA map of Q2237+0305 at L band (20 cm). The off-source map rms is $\sigma=55.6\,\mu$Jy/beam, and the contours are at $(-2,2,4,6,8,10,12,14)\times\sigma$. The crosshair shows the absolute position and uncertainty of the optical galaxy $G$, and the box shows the field displayed in Figure 2. $\Delta\alpha$ and $\Delta\delta$ are referred to a point at $\alpha$=22:37:57.7538, $\delta$=+03:05:49.454 (B1950). The FWHM angular resolution $\Theta \approx 4''$ is shown in the Southeastern corner. The source is unresolved, and the apparent shape is due to antenna phase errors.

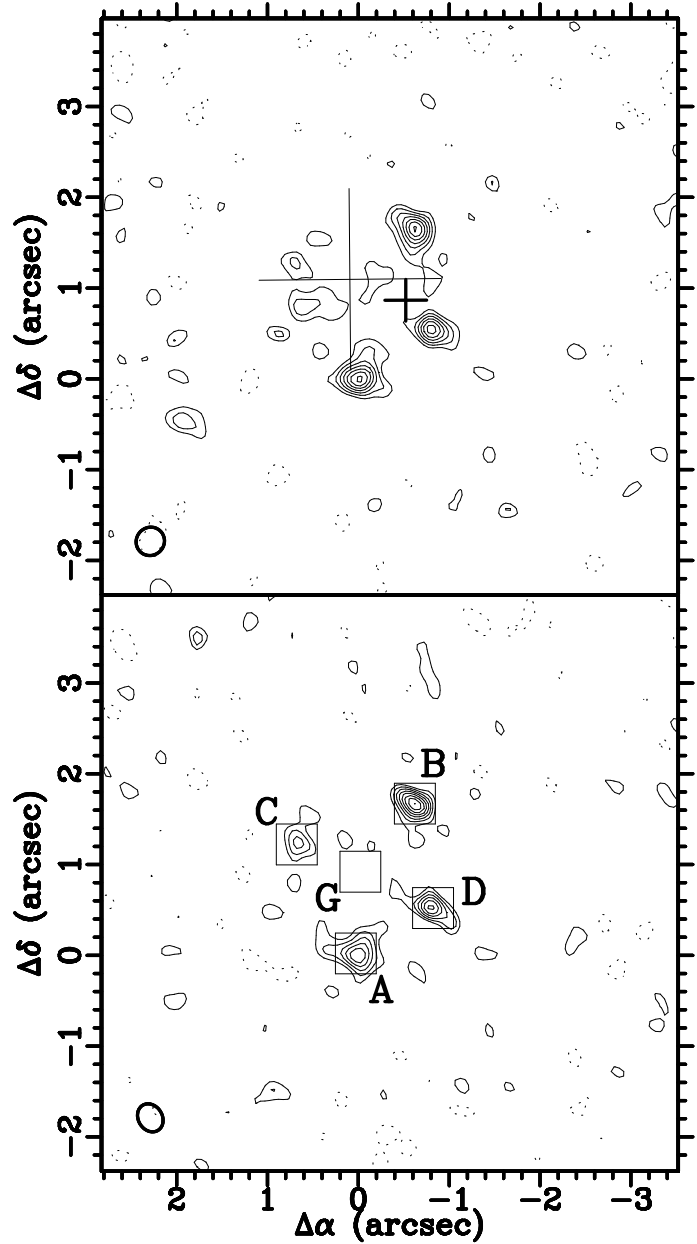

Fig. 2.— VLA maps of Q2237+0305 at X band (3.6 cm). The upper map is from the full observation, with an off-source map rms of $\sigma=7.16\,\mu$Jy/beam. Crosshairs show the absolute positions and uncertainties of the L band source (bold) and the optical galaxy G (light). The lower map is from the restricted data set, with $\sigma=9.82\,\mu$Jy/beam. The photometric apertures are shown as boxes. For both maps, the contours are at $(-2,2,3,4,5,6,7,8,9)\times\sigma$, and positions are referred to $\alpha=22{:}37{:}57.7538$, $\delta=+03{:}05{:}49.454$ (B1950). The FWHM angular resolution $\Theta\approx0\rlap.{''}3$ is shown in the Southeastern corner. The component shapes are due to phase errors.

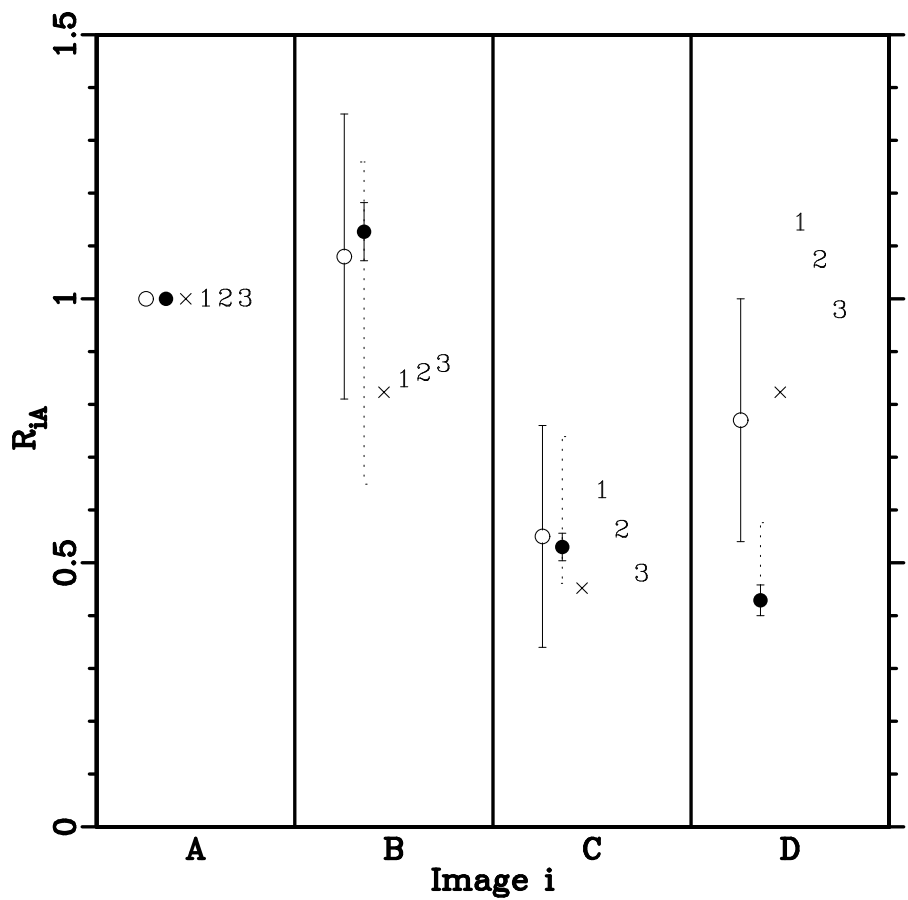

Fig. 3.— Comparison of the observed image brightness ratios $R_{iA}$ with model predictions. The radio ratios (o, this paper) and HST $R$ filter optical ratios (•, Rix et al. 1992) are shown with $1\sigma$ error bars. The dotted lines show the full range of variation in the optical $R$ filter (Houde & Racine 1994). The predicted ratios from various models are also shown: The crosses (×) refer to Model 2a of Rix et al. (1992); and (1,2,3) refer, respectively, to the "best fit", isothermal, and point mass models of Wambsganss & Paczyński (1994).

TABLE 1
Radio Components for Q2237+0305

|   | $\Delta\alpha$ | $\Delta\delta$ | Flux Density | $R_{iA}$ |
|---|---|---|---|---|
| A | $0.000 \pm 0.022$ | $0.000 \pm 0.022$ | $78 \pm 14$ | 1.00 |
| B | $-0.621 \pm 0.017$ | $+1.664 \pm 0.017$ | $85 \pm 14$ | $1.08 \pm 0.27$ |
| C | $+0.664 \pm 0.034$ | $+1.240 \pm 0.034$ | $43 \pm 14$ | $0.55 \pm 0.21$ |
| D | $-0.801 \pm 0.021$ | $+0.530 \pm 0.021$ | $60 \pm 14$ | $0.77 \pm 0.23$ |

All data are from the X band restricted map. Component positions ($\Delta\alpha$,$\Delta\delta$) are in arcseconds, relative to a reference point at $\alpha$=22:37:57.7538, $\delta$=+03:05:49.454 (B1950), with an astrometric error of $< 0\farcs2$. $R_{iA}$ are the component flux ratios (i=B,C,D), relative to A. The flux densities were integrated in $0\farcs45$ square apertures, and are given in $\mu$Jy with $\sim 1\sigma$ uncertainties. The positions were determined using paraboloidal fits to the peaks, and the $\sim 1\sigma$ uncertainties are $\approx \Theta/2Q$ (Fomalont 1989), where $\Theta$ is the FWHM resolution, and $Q$ is the peak signal/noise